\documentclass[%
 aip,
 amsmath,amssymb,
reprint,%
]{revtex4-1}
\usepackage{graphicx}
\usepackage{dcolumn}
\usepackage{lipsum}
\usepackage{bm}
\usepackage{tabu}
\usepackage[utf8]{inputenc}
\usepackage[T1]{fontenc}
\usepackage{mathptmx}
\usepackage{color}

\newcommand{\bl}[1]{\textcolor{blue}{ #1}}
\usepackage{tikz}
\newcommand\encircle[1]{%
  \tikz[baseline=(X.base)] 
    \node (X) [draw, shape=circle, inner sep=-.5] {\strut #1};}
\begin{document}
\preprint{AIP/123-QED}
\title[Measuring the non-separability of vector modes with digital micromirror devices]{\textbf{Measuring the non-separability of vector modes with digital micromirror devices}\\}
\author{Bo-Zhao}
\thanks{This authors contributed equally to this work}
\affiliation{Wang Da-Heng Collaborative Innovation Center for Quantum manipulation \& Control, Harbin University of Science and Technology, Harbin 150080, China}%
\author{Xiao-Bo Hu}
\thanks{This authors contributed equally to this work}
\affiliation{Wang Da-Heng Collaborative Innovation Center for Quantum manipulation \& Control, Harbin University of Science and Technology, Harbin 150080, China}%
\author{Valeria Rodr\'iguez-Fajardo}
\author{Andrew Forbes}
\affiliation{School of Physics, University of the Witwatersrand, Private Bag 3, Johannesburg 2050, South Africa}
\author{Wei Gao}
\author{Zhi-Han Zhu}
\author{Carmelo Rosales-Guzm\'an}
\email{carmelorosalesg@hrbust.edu.cn} 
\affiliation{Wang Da-Heng Collaborative Innovation Center for Quantum manipulation \& Control, Harbin University of Science and Technology, Harbin 150080, China}

\date{\today}
\begin{abstract}
The non-separability between the spatial and polarisation Degrees of Freedom (DoFs) of complex vector light fields has drawn significant attention in recent time. Key to this are its remarkable similarities with quantum entanglement, with quantum-like effects observed at the classical level. Crucially, this parallelism enables the use of quantum tools to quantify the coupling between the spatial and polarisation DoFs, usually implemented with polarisation-dependent spatial light modulators, which requires the splitting of the vector mode into two orthogonal polarisation components. Here we put forward a novel approach that relies on the use of Digital Micromirror Devices (DMDs) for fast, cheap and robust measurement, while the polarisation-independent nature of DMDs enables a reduction in the number of required measurements by 25\%. We tested our approach experimentally on cylindrical vector modes with arbitrary degrees of non-separability, of great relevance in a wide variety of applications. Our technique provides a reliable way to measure in real time the purity of vector modes, paving the way to novel applications where the degree of non-separability can be used as an optical sensor. 

\end{abstract}
\maketitle

\noindent Since their inception in the early 1970s \cite{Dieter1972,Mushiake1972}, complex vector light beams have attracted an increasing amount of interest, in part due to their wide variety of applications \cite{Roadmap,Rosales2018Review,Ndagano2018,Hu2019,Bhebhe2018,BergJohansen2015} but also due to the remarkable similarities these fields share with quantum entangled states \cite{Spreeuw1998,ChavezCerda2007,Qian2011,Aiello2015,konrad2019,forbes2019classically}. More specifically, in complex vector light modes, the spatial and polarisation degrees of freedom (DoFs) are coupled in a non-separable way, akin to the local quantum entanglement observed in a bipartite system, a property that has earned them the name classically entangled modes. Despite the ongoing debate as to whether or not they should be called classically entangled \cite{Karimi2015}, vector modes have been the subject of several studies evincing the parallelism between classical and local quantum entanglement \cite{Balthazar2016,Silva2016,Eberly2016,Toppel2014,Li2016,Ndagano2017}. Given the increasing interest in vector modes, the variety of generation techniques has evolved quite dramatically over the past two decades, amongst which digital holography stands out as one of the most flexible and versatile \cite{Moreno2012,Mitchell2017,SPIEbook,Rosales2017,Ren2015,Rong2014}. While the number of generation techniques evolves continuously, characterisation techniques are still in their infancy. In this way, most techniques to characterise vector modes rely on Stokes polarimetry aiming at the reconstruction of the entire polarisation distribution across the transverse plane \cite{Wen2015,Zhao2019,Rubin2018,Fridman2010}. These techniques, although very powerful, are only qualitative in nature and do not provide quantitative information on the degree of coupling between the spatial and polarisation DoFs. Remarkably, the similarities between classical and quantum entanglement allows the use of the concurrence, $C$, a quantity commonly employed in quantum mechanics to quantify the degree of entanglement of bipartite systems, to measure the non-separability of vector modes\cite{McLaren2015,Ndagano2016,Otte2018,Bhebhe2018a}. This quantity, which has been termed the  Vector Quality Factor (VQF), assigns 0 to a null degree of coupling (scalar modes), and 1 to a maximum degree of coupling (vector modes). To determine the VQF, the vector mode is first projected onto one DoF, let's say polarisation, which is then measured through a series of phase filters by a projection on the spatial DoF. The VQF is then computed from twelve measurements of the on-axis far field intensity. Experimentally, such measurement has been achieved with the help of liquid crystal SLMs or q-plates \cite{Otte2018,Bhebhe2018a,Toninelli2019}. Importantly, SLMs enable the simultaneous measurement of all intensities to perform real-time measurement of the VQF. Nonetheless one of their main disadvantages is their polarisation dependence, which only allows the spatial modulation of one linear polarisation component (typically the horizontal). Hence, in order to measure the VQF of a given vector mode, it has become common to project first over the polarisation DoF, by splitting the vector mode into its two polarisation components, and project afterwards over the spatial DoF encoded as phase patterns on the SLM. 

In this manuscript, we put forward an alternative and more affordable method to measure the concurrence of vector modes. Our method replaces the use of SLMs with digital micromirror devices (DMDs) offering additional advantages, such as, high refresh rates (up to $\sim$ 30 kHz), polarisation independence, wide range of wavelength operation, and low cost. The polarisation independence is crucial as it allows direct spatial projection of the vector mode into all components by the DMD, followed by polarisation DoF projections, e.g., using wave plates and polarisers. Crucially, the change of order of these two operations enables the reduction of the number of measurements by 25 \%, from twelve to eight, and a ``single shot'' measurement operation. To test our technique we used the well-known set of Cylindrical Vector (CV) modes, which are common to many applications, finding a perfect match between experimental measurements and theoretical predictions.

To start with, let recall that the degree of non-separability or "vectorness" of complex light beams can be measured through the Vector Quality Factor (VQF) defined as\cite{Ndagano2016},

\begin{equation}
\text{VQF} =\text{Re}(C)=\text{Re}\left(\sqrt{1-s^2}\right)=|\sin(2\theta)|,
\label{eq:VQF}
\end{equation} 
where {\it C} is the degree of concurrence as defined by Wooters \cite{Wootters1998} and $s$ is the length of the Bloch vector given by,
\begin{equation}
s = \left(\sum_{i=1}^3 \langle\sigma_i\rangle^2\right)^{1/2}.
\label{eq:Blochvector}
\end{equation}
Here, $\langle\sigma_1\rangle$, $\langle\sigma_2\rangle$ and $\langle\sigma_3\rangle$ are the expectation values of the Pauli operators, which represent a set of normalised intensity measurements \cite{McLaren2015}. Physically, {\it s} quantifies the averaged polarisation of the vector modes and can be easily computed from 12 normalised on-axis measurements of the far field intensity. Typically the VQF is performed by splitting the vector mode into its two orthogonal polarisation components, each of which is then projected onto six spatial modes representing the spatial degree of freedom, two orthogonal modes and four additional combinations of the same. Nonetheless, the VQF can be measured by projecting first over two orthogonal spatial modes followed by their projection onto six states of polarisation. For example, considering a vector mode in the circular polarisation basis, such projections would be done onto the left-right, horizontal-vertical and diagonal-antidiagonal polarisation basis. As we will show later, this reduces the number of required measurements from twelve to eight, crucial for the real-time measurement of vector modes with a time-varying degree of non-separability. Importantly, DMDs are the perfect tool to perform such measurements since they enable the modulation of the spatial DoF regardless of its state of polarisation. 

Without any loss of generality, and to use an instructive and topical example, we will restrict our analysis to the well-known Cylindrical Vector (CV) modes given by a non-separable superposition of the Spin and Orbital Angular Momentum (SAM and OAM). With the aim of emphasising the parallelism between quantum and classical entanglement, we will adopt a quantum mechanical notation to express the CV modes. Hence, such a vector mode can be written using the bra-ket notation as, 
\begin{equation}
|\psi \rangle = \cos\left(\theta\right) |+ \ell \rangle | R \rangle +\exp \left(i2\alpha\right) \sin\left(\theta\right) | -\ell \rangle | L \rangle.
\label{VectModes}
\end{equation}
In the above, $| \pm \ell \rangle$ represents the spatial degree of freedom in the OAM basis, where $\ell \in \mathbb{Z}$ is associated to the amount of OAM, $\ell\hbar$ per photon, and $| R \rangle$ and $| L \rangle$ represent the polarisation degree of freedom in the right-left polarisation basis. Further, the amplitude parameter $\theta \in [0, \pi/2]$ allows a continuous tuning of its concurrence, from scalar ($\theta=0$ and $\theta=\pi/2$) to vector ($\theta=\pi/4$). The exponential term $\exp(i2\alpha)$, $\alpha\in [0,\pi]$, is related to the phase difference between the polarisation states.
\begin{figure}[t]
    \centering
    \includegraphics[width=.48\textwidth]{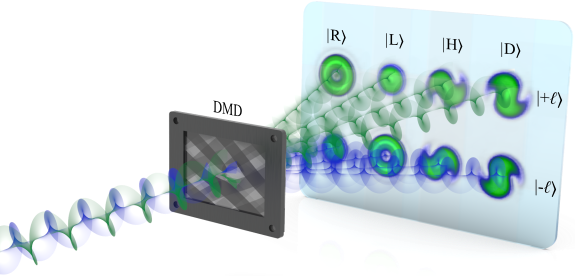}
    \caption{The VQF of vector modes is measured by projecting onto the OAM basis encoded on a DMD and tracing over the polarisation degree of freedom. For this, eight intensity measurements are required, as shown in Table \ref{Tomographic}. Importantly, all intensities can be measured simultaneously using a multiplexing approach.}
    \label{concept}
\end{figure}

As mentioned above, our approach to measure the VQF consists of projecting a given vector mode directly on the $|\pm \ell \rangle$ OAM basis, encoded as binary holograms on a DMD. The resulting mode is then passed through a series of polarisation filters that projects the beam onto the polarisation DoF. Notably, this enables a reduction in the number of required measurements from twelve to eight (see \bl{ Supplementary Material}). This, in combination with a multiplexing approach, facilitates the simultaneous measurement of all the required intensities in a single shot. For the sake of clarity, this is conceptually illustrated in Fig.~\ref{concept}, in which a vector mode is projected onto a digital hologram displayed on the DMD. The hologram consist of a series of eight multiplexed holograms, each with unique spatial frequency, so that the outcomes are directed along independent trajectories.  In this way, four holograms perform the $| +\ell \rangle$ projections and the other four the $| -\ell \rangle$ projections. Thereafter, each beams is passed through a series of polarisation filters to perform the polarisation, namely, onto the $| R \rangle$, $| L \rangle$, $| H \rangle$ and $| D \rangle$ polarisation components. All beams are then passed thought a lens to obtain the far field of each beam, and the on-axis intensity values are then measured to compute the expectation values $\langle\sigma_1\rangle$, $\langle\sigma_2\rangle$ and $\langle\sigma_3\rangle$, from which the VQF can be finally computed. For the sake of clarity, the required projections are shown in Table \ref{Tomographic}. Here, for example, $I_{H\ell^+}$ represents the intensity after projecting the vector mode on the $| +\ell \rangle$ OAM phase filter and passing it through a $| H \rangle$ polarisation filter. 
\begin{table}[b] 
\setlength{\tabcolsep}{10pt}
\renewcommand*{\arraystretch}{2}
 \caption{Normalised intensity measurements $I_{mn}$ to determine the expectation values $\langle \sigma_i \rangle$. \label{Tomographic}}
 \begin{tabular}{c|c c c c c }
Basis states & $| R \rangle$&$| L \rangle$&$| H \rangle$&$| D \rangle$&\\ 
\hline \hline
$| +\ell \rangle$& $I_{R\ell^+}$&$I_{L \ell^+}$&$I_{H\ell^+}$&$I_{D \ell^+}$&\\
$| -\ell \rangle$& $I_{R\ell^-}$ & $I_{L \ell^-}$ & $I_{H\ell^-}$ & $I_{D \ell^-}$
 \end{tabular}
 \end{table}
Explicitly, the expectation values $\langle\sigma_i\rangle$ will take the form (\bl{ Supplementary Material}), 
\begin{align}
\centering
\nonumber
   &\left \langle \sigma_1 \right \rangle=2(I_{H \ell^+}+I_{H \ell^-})-(I_{\ell^+}+I_{\ell^-}),\\
   &\left \langle \sigma_2 \right \rangle=2(I_{D \ell^+}+I_{D \ell^-})-(I_{\ell^+}+I_{\ell^-}),\label{eq:Pauli}\\
    \nonumber
    &\left \langle \sigma_3 \right \rangle=2(I_{R \ell^+}+I_{R \ell^-})-(I_{\ell^+}+I_{\ell^-}),
\end{align}
where, $I_{\ell^+}=I_{R \ell^+}+I_{L \ell^+}=I_{H \ell^+}+I_{V \ell^+}=I_{A \ell^+}+I_{D \ell^+}$ and $I_{\ell^-}=I_{R \ell^-}+I_{L \ell^-}=I_{H \ell^-}+I_{V \ell^-}=I_{A \ell^-}+I_{D \ell^-}$.
\begin{figure}[tb]
    \centering
    \includegraphics[width=0.49\textwidth]{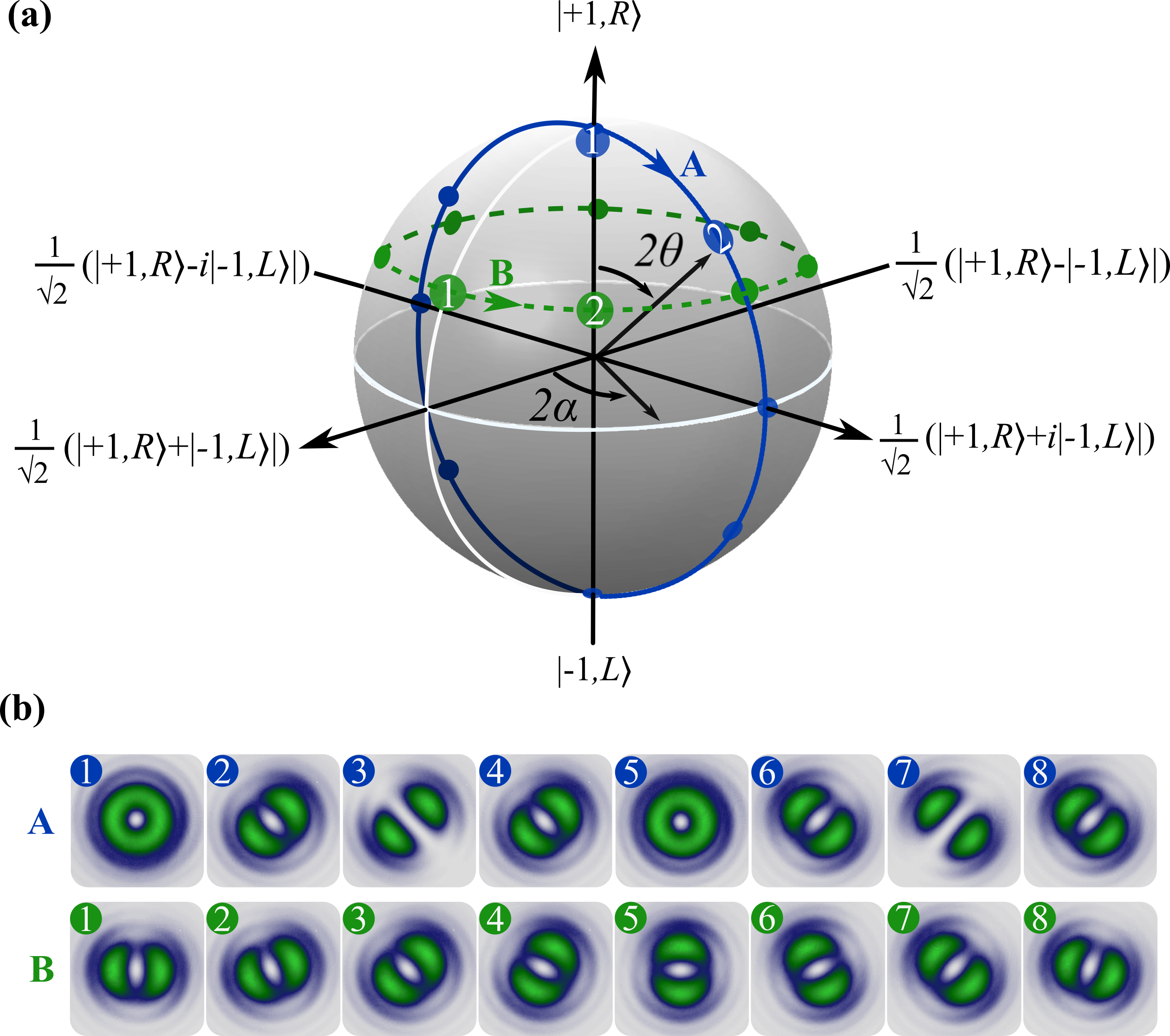}
    \caption{(a) Geometric representation of vector modes using the HOPS. The case $\alpha=\pi/4$, $\theta \in [0,\pi]$ generates modes along path {\bf A} (solid line), which are shown in the top row of (b). The case $\theta=\pi/6$, $\alpha\in[0,\pi]$ generates the modes along path {\bf B} (dashed line ), which are shown in the bottom row. The numbers indicates the position of each beam on the HOPS.}
    \label{sphere}
\end{figure}

To demonstrate our technique, we prepared vector modes with arbitrary degrees of concurrence using a liquid crystal q-plate ($q=1/2$) in combination with a quarter- and a half- wave plate \cite{Marrucci2006,Karimi2010}. The generated vector modes are shown in Fig.~\ref{sphere}, first, in Fig.~\ref{sphere}(a) we show their geometrical representation as points $(\alpha, \theta)$ on the surface of the so-called High Order Poincare Sphere (HOPS). Here, points along path {\bf A} (blue solid line) connecting the north and south poles represent vector states with varying degrees of concurrence, obtained by varying $\theta$ in the interval [0,$\pi$] while maintaining $2\alpha =\pi/2$. The intensity profile of such states, after passing a linear polariser, is shown in the top row of Fig.~\ref{sphere}(b). Notice how the beam changes from scalar (Fig. \ref{sphere}(b)-1) to vector (Fig.~\ref{sphere}(b)-3) and then back to scalar (Fig. \ref{sphere}(b)-5) and back to vector again (Fig.~\ref{sphere}(b)-7). We also generated the vector modes represented along the path parallel to the equator (green dashed line) with a constant degree of concurrence but varying intermodal phase $2\alpha\in[0,2\pi]$. Their intensity profile after transmitted through a linear polariser is shown in the bottom row of Fig.~\ref{sphere}(b). Notice how the intensity distribution rotates as $\alpha$ increases, maintaining the same intensity shape.

\begin{figure}[tb]
\includegraphics[width=.48\textwidth]{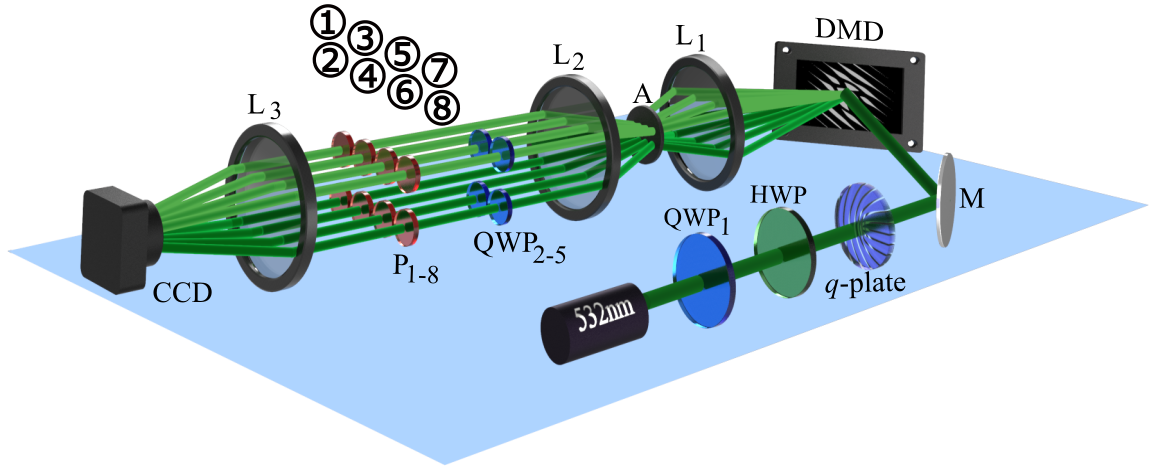}
\caption{\label{setup} Vector modes with arbitrary degrees of concurrence are generated from a CW Gaussian beam ($\lambda=532$ nm) using a liquid crystal q-plate (q=1/2) in combination with a Quarter- and a Half-Wave Plate (QWP1). Afterwards, a Digital Micromirror Device (DMD) performs its projection onto the OAM modal basis. A multiplexing approach enables the simultaneous generation of 8 beams, the top four carrying the $|+\ell \rangle$ and the bottom four carrying the $|-\ell \rangle$ projection, which are then passed through a series of optical filters to trace over the polarisation degree of freedom. Finally, lens L3 focuses all beams onto a CCD camera.}
\end{figure}

Figure \ref{setup} illustrates the experimental setup implemented to measure the VQF. Our vector beams were generated from a linearly polarised Gaussian beam ($\lambda=532$ nm), using a liquid crystal q-plate ($\ell=1$ WPV10L-532, Thorlabs). The generated modes were first directed onto a multiplexed binary hologram displayed on a DMD (DLP Light Crafter 6500 from Texas Instruments) that performs the projection onto the spatial DoF. The hologram consisted of eight multiplexed holograms with unique carrier frequencies, four identical holograms that perform the projection $|+\ell \rangle$ (top row) and another four that perform the projection $|-\ell \rangle$ (bottom row). The carrier frequencies are carefully chosen to separate the beams along eight independent trajectories, which are then spatially filtered, to remove higher diffraction orders, and collimated using lenses L$_1$ and L$_2$ ($f_{1,2}=200$ mm) to propagate along parallel paths. Each of these beams are then transmitted through polarising optical filters to acquire the required intensities $I_{mn}$ as explained next. Beams \encircle{1} and \encircle{2} are transmitted through a linear polariser at $0^\circ$ to obtain $I_{H \ell^+}$ and $I_{H \ell^-}$, respectively. Beams \encircle{3} and \encircle{4} are passed through a linear polariser oriented at $45^\circ$ to measure $I_{D \ell^+}$ and $I_{D \ell^-}$, respectively. Beams \encircle{5} and \encircle{6} are used to measure $I_{R \ell^+}$ and $I_{R \ell^-}$, by transmitting them through a Quarter Wave-plate (QWP) at $45^o$ and a linear polariser at $90^\circ$. Finally, beams \encircle{7} and \encircle{8} are used to measure $I_{L \ell^+}$ and $I_{L \ell^-}$ by means of a QWP at $-45^\circ$ and a linear polariser at $90^\circ$. Finally, the far field of all beams is acquired using a third lens (L$_3$, $f_3$=200mm), which focuses the beams onto a CCD camera (BC106N-VIS Thorlabs) that records all the intensities simultaneously. 

\begin{figure}[tb]
    \centering
    \includegraphics[width=0.48\textwidth]{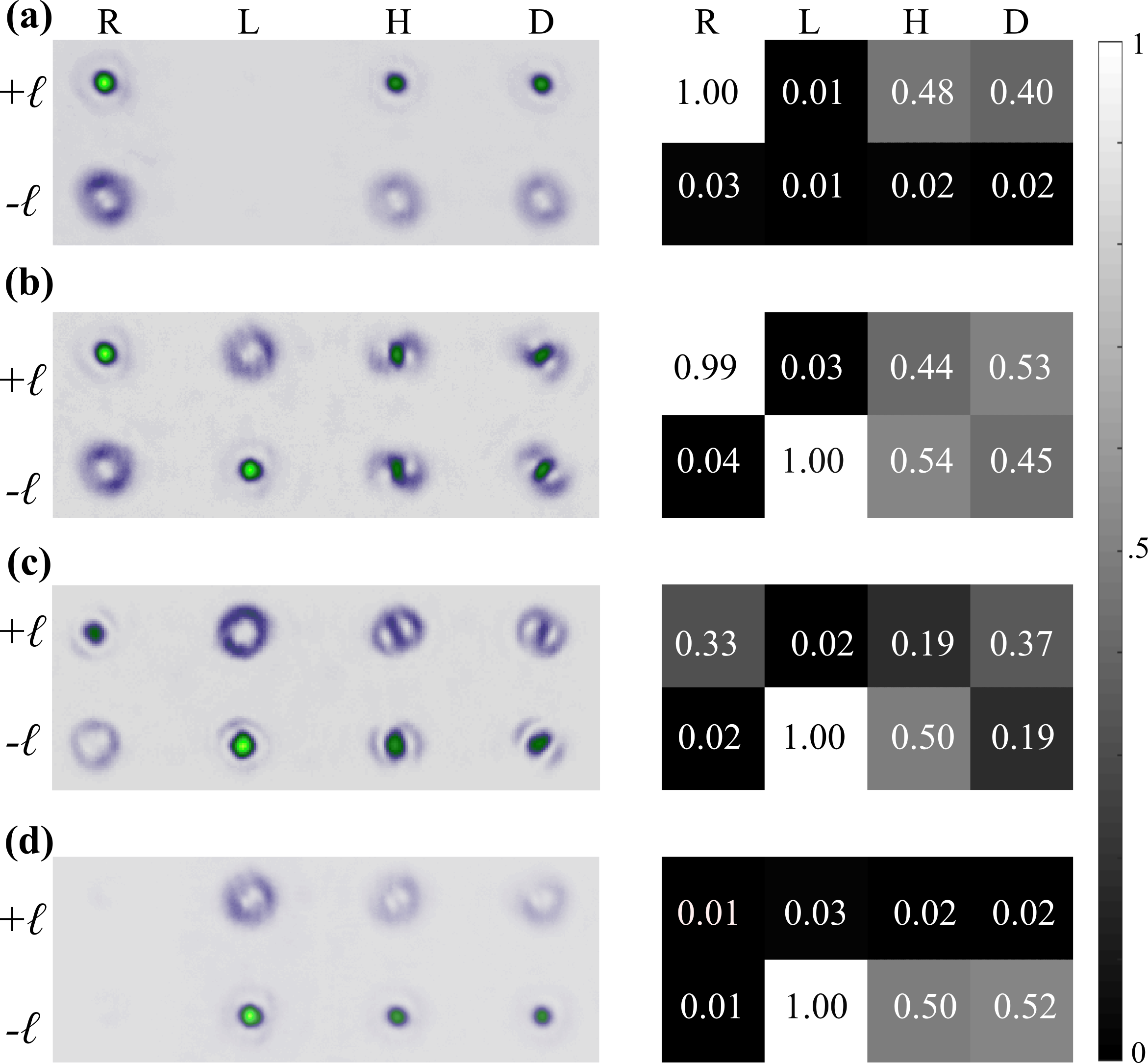}
    \caption{Example images of the intensity recorded on the CCD (left column) and their on-axis intensity values (right column) for four different vector modes. (a) Scalar mode with right circular polarisation and VQF=0.0, (b) Pure vector mode with VQF=0.98, (c) non-pure vector mode with VQF=0.70 and (d) Scalar mode with left circular polarisation and VQF=0.00.}
    \label{Example}
\end{figure}

Figure \ref{Example} shows four examples of vector modes with different degrees of VQF to further illustrate our measurement procedure. First, in Fig. \ref{Example}(a), we show the case $\theta=0$, which corresponds to the state $|\Psi \rangle=|+ \ell \rangle | R \rangle$ (see Eq. \ref{VectModes}). On the left we show the intensity images as recorded by the CCD, while on the right we show the intensity values, normalised to the maximum intensity, along the optical axis. Notice that the only nonzero values are obtained for the projections on the $|+\ell \rangle$ state for the $|R \rangle$, $|H \rangle$ and $|D \rangle$ polarisation, acquiring values 1.00, 0.48 and 0.40, respectively. Substitution of these intensity values into Eqs.\ref{eq:Pauli} and \ref{eq:VQF} yields a value VQF=0.00, which indeed correspond to a scalar mode. Figure \ref{Example}(b) shows the case $\theta=\pi/4$ and $\alpha=\pi/2$, which give rise to a pure vector mode. For this case we get maximum intensity values at $|+\ell \rangle|R \rangle$ and $|-\ell \rangle|L \rangle$, 0.99 and 1.00, respectively. Further, we also get nonzero values at the four projection given by $|\pm\ell \rangle|H \rangle$ and $|\pm\ell \rangle|D \rangle$, 0.44, 0.53, 0.54 and 0.45, respectively. Again, substitution of these values onto Eq. \ref{eq:Pauli} and \ref{eq:VQF}, respectively, yields the value VQF=0.98. In Fig. \ref{Example}(c) we show the case of a vector mode with an intermediate value of VQF. As can be see, the on-axis intensities varies in a more intricate way, which results in a VQF value of 0.70. Finally, the case $\theta=\pi/2$ is shown in Fig. \ref{Example}(d), which corresponds to the mode $|\Psi \rangle=|- \ell \rangle | L \rangle$. Here, the nonzero intensities are obtained at the $|-\ell \rangle$ projection for the $|L \rangle$, $|H \rangle$ and $|D \rangle$ polarisations, 1.00, 0.50 and 0.52, respectively, which yields the value VQF=0.00 It is worth mentioning that the on-axis intensities can be measured easily with the use of a calibration image that provides the ($x,y$) coordinates of the optical axis of each beam. This procedure is further detailed in \bl{Supplementary Material}.

\begin{figure}[tb]
    \centering
    \includegraphics[width=.45\textwidth]{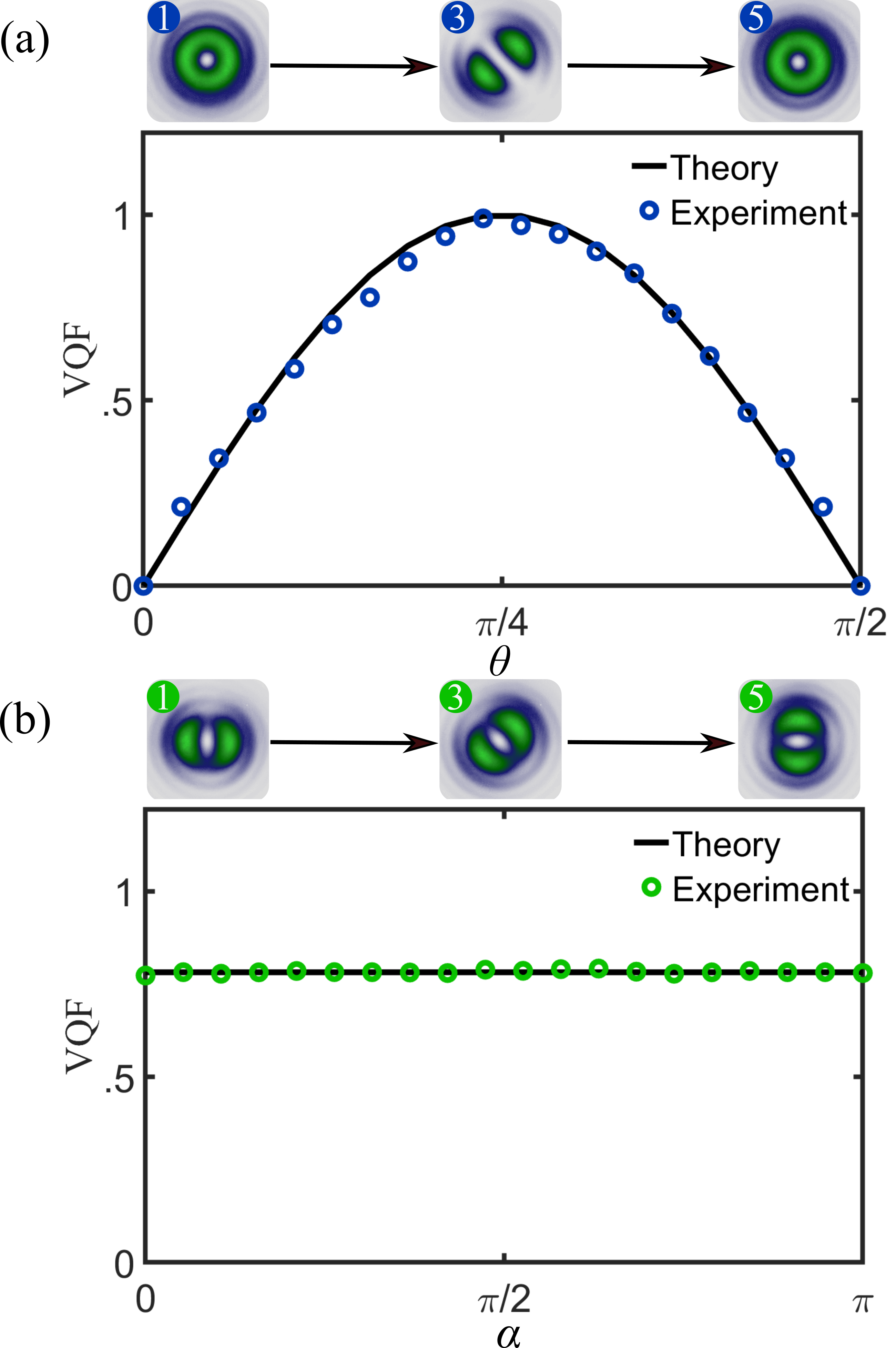}
    \caption{VQF of vector modes as function of the parameters $\theta$ and $\alpha$. The solid line corresponds to theory and the circles to experimental measurements. (a) The case $\alpha=\pi/4$ and $\theta\in[0,\pi/2]$ produces vector modes with varying degrees of non-separability. (b) The case $\theta=\pi/6.$ and $\alpha\in[0,\pi]$ produces modes with a constant degree of non-separability.}
    \label{Results}
\end{figure}

Finally, in Fig. \ref{Results}, we show a plot of the VQF as function of both, $\theta$ and $\alpha$, the theory (Eq. \ref{eq:VQF}) is represented by the continuous line and the experimental data by points. Figure \ref{Results}(a) shows the case where $\theta$ is varied from 0 to $\pi/2$, which corresponds to the states shown in the top row of Fig. \ref{sphere} (b). Notice how the VQF increases continuously from 0 at $\theta=0$ to 1 at $\theta=\pi/4$ and goes back to zero for $\theta=\pi/2$, as predicted by theory. Figure \ref{Results}(b) shows the case $\theta=\pi/6$ and $\alpha\in[0,\pi]$, which corresponds to the states shown in the bottom row of Fig. \ref{sphere} (b). As expected, in this case the VQF remains constant for all values of $\alpha$. Notice the high agreement of our experimental measurements with the theoretical predictions. 

In conclusion, here we put forward a novel technique to measure the degree of non-separability of classically-entangled states of light. Our device takes full advantage of DMD technology, particularly, their polarisation insensitivity. Even though DMDs have been around for decades, it is only in recent time that they started to be used as optical modulators to generate structured light beams, owing to their advantages over commonly used SLMs. Their polarisation insensitivity allows them to modulate any polarisation state, a property that has been exploited to generate arbitrary vector beams \cite{Selyem2019}. This property has been also exploited in the development of a technique capable to reconstruct the transverse polarisation distribution of vector beams in real time \cite{Zhao2019}. In this manuscript we presented a technique that allowed us to project an input vector beam directly onto a DMD to assess its purity. More precisely, we project the mode onto the spatial DoF specified by the OAM basis, which is then projected onto the polarisation DoF using a series of polarisation filters composed of wave plates and polarisers. Importantly, this approach enables the reduction of the number of required measurements by 25\%. Moreover, DMDs are one order of magnitude cheaper compared to SLMs, at least one order of magnitude faster, making our approach ideal for the single shot quantitative analysis of vector modes. 

\section*{Supplementary material}
see \bl{supplementary material} for the derivation of Eq. \ref{eq:Pauli} and for additional details of the on-axis intensity measurements.

\section*{Funding}
 This work was partially supported by the National Nature Science Foundation of China (NSFC) under Grant Nos.  61975047, 11934013, 11574065.
 \section*{References}

\end{document}


\preprint{AIP/123-QED}
\title[Measuring the non-separability of vector modes with digital micromirror devices]{\textbf{Supplementary material for: Measuring the non-separability of vector modes with digital micromirror devices}\\}
\author{Bo-Zhao}
\thanks{This authors contributed equally to this work}
\affiliation{Wang Da-Heng Collaborative Innovation Center for Quantum manipulation \& Control, Harbin University of Science and Technology, Harbin 150080, China}%
\author{Xiao-Bo Hu}
\thanks{This authors contributed equally to this work}
\affiliation{Wang Da-Heng Collaborative Innovation Center for Quantum manipulation \& Control, Harbin University of Science and Technology, Harbin 150080, China}%
\author{Valeria Rodr\'iguez-Fajardo}
\author{Andrew Forbes}
\affiliation{School of Physics, University of the Witwatersrand, Private Bag 3, Johannesburg 2050, South Africa}
\author{Wei Gao}
\author{Zhi-Han Zhu}
\author{Carmelo Rosales-Guzm\'an}
\email{carmelorosalesg@hrbust.edu.cn} 
\affiliation{Wang Da-Heng Collaborative Innovation Center for Quantum manipulation \& Control, Harbin University of Science and Technology, Harbin 150080, China}

\date{\today}
\begin{abstract}
In this supplementary material we provide additional information about the explicit derivation of Eq. 4 of our main manuscript, which allows to reduce the number of required on-axis intensity measurements, from twelve to eight. In addition, here we also provide with additional details about the measurement of all the required on-axis intensity in a single-shot to perform real-time measurement of the VQF.

\end{abstract}
\maketitle

\def\theequation{S\arabic{equation}}
\section*{Determination of VQF from eight intensity measurements}
To compute the degree of non-separability of vector modes we first need to compute the expectation values of the Pauli operators, namely, $\langle\sigma_1\rangle$, $\langle\sigma_2\rangle$ and $\langle\sigma_3\rangle$. In our approach, we compute these by projecting a given vector mode onto the spatial DoF, which is afterwards traced over the polarisation DoF. The $\langle\sigma_i\rangle$ values are then computed from on-axis measurements of the far field intensity. This technique requires the measurement of 12 intensities, explicitly shown in Table \ref{Tomographic1}. Nonetheless, this set of intensities can be reduced to only eight, as will be explained below.  
\begin{table}[b] 
\setlength{\tabcolsep}{7pt}
\renewcommand*{\arraystretch}{2}
 \caption{Normalised intensity measurements $I_{mn}$ to determine the expectation values $\langle \sigma_i \rangle$. \label{Tomographic1}}
 \begin{tabular}{c|c c c c c c c}
Basis states & $| R \rangle$ & $| L \rangle$ & $| H \rangle$ & $| D \rangle$& $| V \rangle$ & $| A \rangle$& \\
\hline \hline
$| +\ell \rangle$& $I_{R\ell^+}$ & $I_{L \ell^+}$ & $I_{H \ell^+}$ & $I_{D \ell^+}$&  $I_{V \ell^+}$ & $I_{A \ell^+}$  &\\ 
$| -\ell \rangle$& $I_{R\ell^-}$ & $I_{L \ell^-}$ & $I_{H \ell^-}$ & $I_{D \ell^-}$& $I_{V \ell^-}$ & $I_{A \ell^-}$& \\
\end{tabular}
\end{table}

The expectation values $\langle\sigma_i\rangle$ are computed as,
\begin{align}
\centering
   \nonumber
   &\left \langle \sigma_1 \right \rangle=(I_{H \ell^+}+I_{H \ell^-})-(I_{V\ell^+}+I_{V\ell^-}),\\
   &\left \langle \sigma_2 \right \rangle=(I_{D \ell^+}+I_{D \ell^-})-(I_{A\ell^+}+I_{A\ell^-}),\label{eq:Pauli2}\\
    &\left \langle \sigma_3 \right \rangle=(I_{R \ell^+}+I_{R \ell^-})-(I_{L\ell^+}+I_{L\ell^-})\nonumber,
\end{align}
To see how this set of measurements can be reduced from twelve to only eight notice that $I_{R \ell^+}+I_{L \ell^+}=I_{H \ell^+}+I_{V \ell^+}=I_{D \ell^+}+I_{A \ell^+}$. Hence, the second term of $\langle\sigma_1\rangle$ in Eq. \ref{eq:Pauli2} can be written as,
\begin{equation}
    \centering
   I_{V\ell^+}+I_{V\ell^-}=I_{V\ell^+}+I_{H\ell^+}-I_{H\ell^+}+I_{V\ell^-}+I_{H\ell^-}-I_{H\ell^-},
    \label{sigma1}
\end{equation}
where we have added and subtracted the same term, that is, $I_{H\ell^+}-I_{H\ell^+}=0$ and $I_{H\ell^-}-I_{H\ell^-}=0$. Rearrangement of the terms in Eq. \ref{sigma1} leads to,
\begin{align}
    \centering
    \nonumber
   I_{V\ell^+}+I_{V\ell^-}&=-(I_{H\ell^+}+I_{H\ell^-})+I_{V\ell^+}+I_{H\ell^+}+I_{V\ell^-}+I_{H\ell^-}.\\
   &=-(I_{H\ell^+}+I_{H\ell^-})+I_{\ell^+}+I_{\ell^-}.
    \label{sigma2}
\end{align}
Where, $I_{\ell^+}=I_{V\ell^+}+I_{H\ell^+}$ is the total intensity of the vector mode projected onto the mode $| +\ell \rangle$ and similarly, $I_{\ell^-}=I_{V\ell^-}+I_{H\ell^-}$ is the total intensity of the projection onto the mode $| -\ell \rangle$. Hence, $\langle\sigma_1\rangle$ becomes,
\begin{equation}
    \centering
   \left \langle \sigma_1 \right \rangle=2(I_{H \ell^+}+I_{H \ell^-})-(I_{H\ell^+}+I_{V\ell^-}).
    \label{sigma1f}
\end{equation}
A similar procedure for $\langle\sigma_2\rangle$ and $\langle\sigma_3\rangle$ leads to the final expression,
\begin{align}
\centering
\nonumber
   &\left \langle \sigma_1 \right \rangle=2(I_{H \ell^+}+I_{H \ell^-})-(I_{\ell^+}+I_{\ell^-}),\\
   &\left \langle \sigma_2 \right \rangle=2(I_{D \ell^+}+I_{D \ell^-})-(I_{\ell^+}+I_{\ell^-}),\\
    \nonumber
    &\left \langle \sigma_3 \right \rangle=2(I_{R \ell^+}+I_{R \ell^-})-(I_{\ell^+}+I_{\ell^-}),
    \label{Pauli3}
\end{align}
which is Eq. 4 of our main manuscript.

\section*{Measurement of the on-axis intensities}
\begin{figure}[tb]
    \centering
    \includegraphics[width=0.48\textwidth]{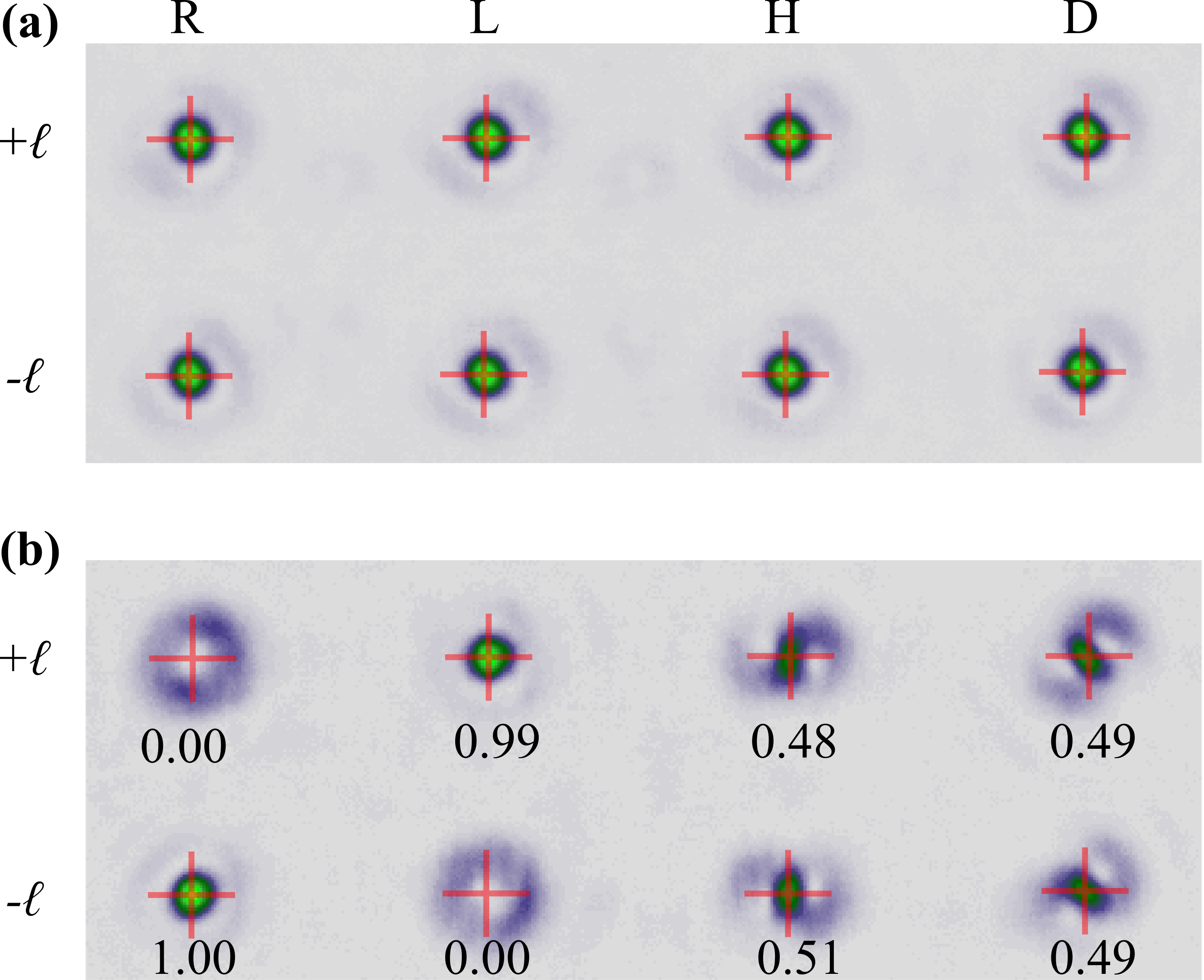}
    \caption{(a) Calibration image to find the on-axis coordinates for each beam. (b) Example image and their intensity values read at the coordinates defined in (a).}
    \label{calibration}
\end{figure}

Given that our technique employees a multiplexing approach to measure all on-axis intensities simultaneously, we first need to define the on-axis coordinates of all the beams. To this end, we use a calibration image consisting of 8 Gaussian modes, from which the on-axis coordinates are extracted by finding the maximum intensity values of each Gaussian mode. Fig. \ref{calibration}(a) shows such calibration image where the coordinates of each maximum intensity, given by the coordinates $(x_n,y_m)$, have been marked with a cross. This step is required only one time, once the $(x_n,y_m)$ coordinates are found, a process that can be done automatically using any programming software, these are saved and can be used as long as the system remains unchanged. Figure \ref{calibration}(b), shows an example of an image produced when a vector beams is passed through our system. Here, and for the sake of clarity, each of the on-axis coordinates where the intensities are measured have been marked with the same cross as in Fig. \ref{calibration}(a). Further, all the intensities are normalised to the maximum intensity value.
